\documentclass[twocolumn]{aastex7}

\received{April 18, 2025}
\revised{June 14, 2025}
\accepted{July 15, 2025}

\begin{document}

\title{Black Hole Survival Guide: \\Searching for Stars in the Galactic Center That Endure Partial Tidal Disruption}

\shorttitle{Stars That Survive Tidal Disruption}
\shortauthors{Bush et al.}

\author[0000-0002-6979-5388,gname='Rewa',sname='Bush']{Rewa Clark Bush}
\affiliation{Department of Astronomy and Astrophysics, University of California, Santa Cruz, CA 95064, USA}
\affiliation{Department of Physics, Cabrillo College, Aptos, CA 95003, USA}
\affiliation{Department of Astronomy and Van Vleck Observatory, Wesleyan University, Middletown, CT 06459, USA}
\email{rbush@wesleyan.edu}

\author[0000-0003-2872-5153,gname='Samantha',sname='Wu']{Samantha C. Wu}
\affiliation{The Observatories of the Carnegie Institution for Science, Pasadena, CA 91101, USA}
\affiliation{Center for Interdisciplinary Exploration \& Research in Astrophysics (CIERA), Physics \& Astronomy, Northwestern University, Evanston, IL 60202, USA}
\email{swu@carnegiescience.edu}

\author[0000-0001-5256-3620,gname='Rosa',sname='Everson']{Rosa Wallace Everson}
\affiliation{Department of Astronomy and Astrophysics, University of California, Santa Cruz, CA 95064, USA}
\affiliation{Department of Physics and Astronomy, University of North Carolina at Chapel Hill, Chapel Hill, NC 27599, USA}
\email{rosa.wallace.everson@gmail.com}

\author[0000-0003-0381-1039,gname='Ricardo',sname='Yarza']{Ricardo~Yarza}
\altaffiliation{NASA FINESST Fellow}
\affiliation{Department of Astronomy and Astrophysics, University of California, Santa Cruz, CA 95064, USA}
\email{ryarza@ucsc.edu}

\author[0000-0003-2333-6116,gname='Ariadna',sname='Murguia-Berthier']{Ariadna Murguia-Berthier}
\altaffiliation{NASA Hubble Postdoctoral Fellow}
\affiliation{Center for Interdisciplinary Exploration \& Research in Astrophysics (CIERA), Physics \& Astronomy, Northwestern University, Evanston, IL 60202, USA}
\affiliation{Department of Astronomy and Astrophysics, University of California, Santa Cruz, CA 95064, USA}
\email{arimurguia@northwestern.edu}

\author[0000-0003-2558-3102,gname='Enrico',sname='Ramirez-Ruiz']{Enrico Ramirez-Ruiz}
\affiliation{Department of Astronomy and Astrophysics, University of California, Santa Cruz, CA 95064, USA}
\email{raruiz@ucsc.edu}

\correspondingauthor{Rewa Clark Bush}
\email{rbush@wesleyan.edu}

\begin{abstract}

Once per $\approx10^4$--$10^5$ years, an unlucky star may experience a close encounter with a supermassive black hole (SMBH), partially or fully tearing apart the star in an exceedingly brief, bright interaction called a tidal disruption event (TDE). Remnants of partial TDEs are expected to be plentiful in our Galactic Center, where at least six unexplained, diffuse, star-like ``G objects'' have already been detected which may have formed via interactions between stars and the SMBH. Using numerical simulations, this work aims to identify the characteristics of TDE remnants. We take 3D hydrodynamic FLASH models of partially disrupted stars and map them into the 1D stellar evolution code MESA to examine the properties of these remnants from tens to billions of years after the TDE. The remnants initially exhibit a brief, highly luminous phase, followed by an extended cooling period as they return to stable hydrogen burning. During the initial stage ($\lesssim10^5$ yr) their luminosities increase by orders of magnitude, making them intriguing candidates to explain a fraction of the mysterious G objects. Notably, mild TDEs are the most common and result in the brightest remnants during this initial phase. However, most remnants exist in a long-lived stage where they are only modestly offset in temperature and luminosity compared to main-sequence stars of equivalent mass. Nonetheless, our results indicate remnants will sustain abnormal, metal-enriched envelopes that may be discernible through spectroscopic analysis. Identifying TDE survivors within the Milky Way could further illuminate some of the most gravitationally intense encounters in the Universe.

\end{abstract}


\section{Introduction} \label{sec:intro}

Galactic nuclei are natural sites for both collisions and tidal interactions of stars, given the high concentration of stars in the proximity of the central supermassive black hole (SMBH). The extreme stellar density and high velocity dispersion in galactic nuclei \citep{Genzel2003, Ghez2003, Ferrarese2005, Schodel2018} may give rise to the formation of unusual collisional \citep{Gibson2025, Stephan2016} and tidally stripped stellar remnants \citep{Guillochon2014, DeColle2014}, such as  exotic massive stars.

Specifically, peculiar dust- and gas-enshrouded stellar objects, called G objects, have been observed in the Milky Way’s nuclear star cluster \citep{Gillessen2012, Witzel2014}: six known G objects are presented in \citet{Ciurlo2020}. Their origins remain unknown, although several formation channels have been proposed in the literature \citep[e.g.,][]{Burkert2012, Schartmann2012, Guillochon2014, Madigan2017, Owen2023}. The proposed formation channels include, for example, stellar mergers \citep[e.g.,][]{Witzel2014, Prodan2015, Stephan2016, Stephan2019, Wu2020} and stellar collisions \citep[e.g.,][]{Rose2023}. Here we investigate the viability of a third formation avenue: the puffy stellar remnant formed when a star survives a tidal encounter with the central SMBH.

At the center of the Milky Way, the gravitational landscape is dominated by Sgr A*, a SMBH with a mass roughly four million times that of our Sun \citep[e.g.,][]{Ghez2008}. Approximately 10 million stars swarm this SMBH in tightly packed orbits within a few parsecs of the Galactic Center, an environment that is $\lesssim 10^8$ times denser than our Solar neighborhood \citep[e.g.,][]{Lu2013, Schodel2014, Schodel2018}. Each star within this region traces out an intricate orbit under the combined influence of the SMBH and all other stars. The tidal radius $R_\tau \equiv R_\ast (M_{\rm BH}/M_\ast)^{1/3}$ is defined as the boundary of closest approach, beyond which point the SMBH's gravity overcomes the star's self gravity and the SMBH begins to unbind the star and siphon stellar mass. Here $M_{\rm BH}$ is the mass of the SMBH and  $M_\ast$ and $R_\ast$ are the stellar mass and radius, respectively. If a star wanders past this critical distance, a tidal disruption event (TDE) occurs, as the star is fiercely ripped apart by the SMBH’s tidal field \citep[e.g.,][]{Hills1975, Rees1988, Alexander1999, Magorrian1999, Wang2004}.

TDEs are important to study because they can illuminate previously quiescent SMBHs, revealing key properties about those SMBHs in the process and ultimately contributing to population demographics \citep[e.g.,][]{Guillochon2013, Yang2017, Dai2021, Mockler2022, Miller2023}. Tidal encounters between SMBHs and stars are also phenomenal opportunities to deepen our understanding of environments with intense gravity \citep[e.g.,][]{Guillochon2009}. However, TDEs are rare events. The prevalence of TDEs is heavily dependent on cosmic epoch, merger history, and galaxy properties, but a conservative estimate suggests an occurrence rate of approximately $10^{-4}$ to $10^{-5}$ TDEs per galaxy per year \citep{Alexander1999,Magorrian1999, Wang2004, Stone2016, Pfister2021}. In other words, to guarantee seeing a TDE in progress would require staring at the nucleus of the Milky Way for 100,000 years, or alternatively, searching 100,000 galaxies every day for a year. Most have chosen the latter route, utilizing large surveys of galaxies in hopes of chancing upon the transient flare of a TDE in a galaxy far, far away. At the time of this paper's publication, fewer than one hundred candidate TDEs have been observed using this strategy.\footnote{\url{https://github.com/astrocatalogs/tidaldisruptions}}

In this {\it Letter}, we aim to ascertain what the remnants of these tidal encounters look like in our own backyard -- the Galactic Center of the Milky Way. A significant portion of TDEs do not result in total annihilation of the star, leaving behind a warped stellar remnant that survives and is ejected from the BH \citep{Stone2016, Krolik2020}. These are known as ``partial TDEs,’’ and are estimated to occur at a rate 10$\times$ higher than full TDEs \citep{Bortolas2023}. In the 13.6-billion-year history of the Milky Way, for a partial TDE occurrence rate of $\sim 10^{-4}\, \mathrm{yr}^{-1}$, there should have been time for approximately one million such ``survivor stars'' to accumulate in our midst. These remnant stars could bring insights into some of the most intense encounters with extreme gravity in the Universe. Here we investigate how TDE survivors evolve, from tens to billions of years after disruption, and we seek the fingerprints that will differentiate them from surrounding stars. Our research combines hydrodynamical and stellar evolutionary simulations to identify whether TDE survivors retain any signature characteristics that could clue us into their pasts and facilitate their detection. 

\begin{figure*}[ht!]
\includegraphics[width=\textwidth]{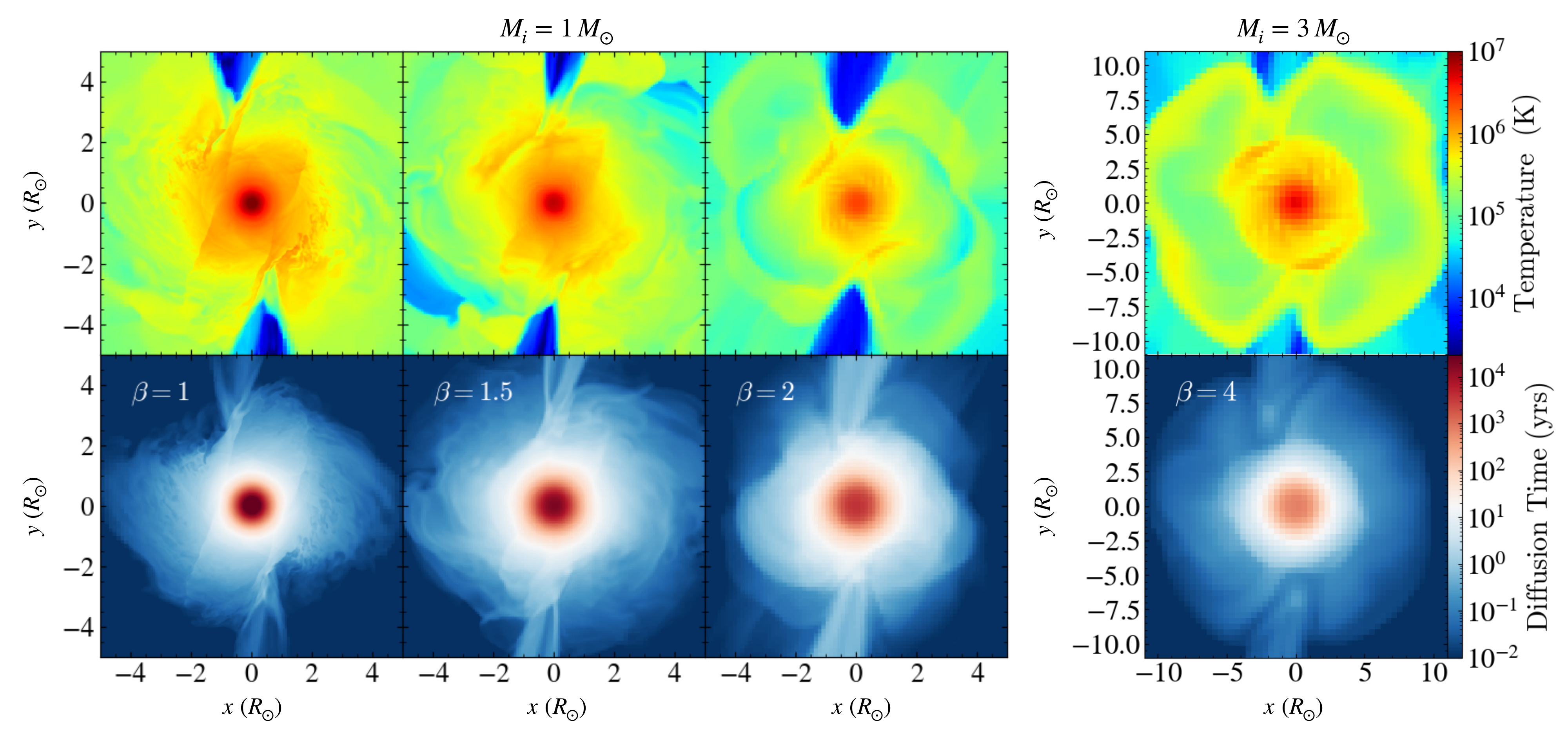}
\caption{Slices in 2D of the four TDE remnants studied in this paper, selected from the STARS Library \citep{LawSmith2020}. The left three columns show remnants that begin as a 1 $M_\odot$, 4.8 Gyr main sequence star before disruption, and the fourth displayed in the rightmost column begins as a 3 $M_\odot$, 0.3 Gyr main sequence star. 
Each is subjected to varying degrees of tidal disruption. The impact parameter ($\beta$) increases from left to right, indicating how deeply the star plunged past the SMBH's tidal radius. Colorbars indicate temperature in Kelvin (top row) and diffusion time in years (bottom row). The left panels have width 10 $R_{\odot}$, while the rightmost column has width 22 $R_{\odot}$. All snapshots are captured at roughly 80 $t_{\rm dyn}$ after pericenter, corresponding to $\approx 38$ hr after pericenter for the left three columns and $\approx 124$ hr after pericenter for the right column.}
\label{fig:hydro}
\end{figure*}

To account for the distorted internal structure of the star during a TDE, a hydrodynamical description must be employed to effectively describe the remnant's deformation and the exchange of energy and angular momentum during the encounter \citep{Evans1989, Ivanov2001, Khokhlov1993, Laguna1993, Ramirez-Ruiz2009, MacLeod2016, LawSmith2020}. Then a new stellar model may be constructed for the remnant, often with a highly unusual chemical composition and physical conditions. The timescale for the remnant to regain its thermal equilibrium is vastly longer than the minutes or days needed for dynamical equilibrium to be restored, which warrants pivoting to one-dimensional stellar evolution simulations in order to capture the further evolution \citep[e.g.,][]{Schneider2020, Wu2020}.

In our study, we utilize the open-source 3D hydrodynamical code FLASH \citep{Fryxell2000} to simulate the TDE, then pivot to the 1D stellar evolution code MESA \citep{Paxton2011, Paxton2013, Paxton2015, Paxton2018} to capture the TDE survivor's subsequent evolution.
Our results at late times are consistent with recent work investigating the long-term evolution ($>100$ Myr) of TDE survivors from \cite{Sharma2024}. What sets our study apart is that our methods enabled us to investigate the initial ($>1000$ yr) bright phase of the remnant as well, and thereby quantify the remnant's behavior during what may be the most opportune period for their detection as unusual stellar objects.

\section{Numerical Framework} \label{sec:initial}

We aim to capture the details of the tidal disruption, which takes place over a few dynamical timescales of the star, as well as the remnant's subsequent long-term evolution, lasting upwards of a few billion years. To address this challenge, we have combined simulations of TDEs using the adaptive mesh hydrodynamic code FLASH \citep{Fryxell2000} paired with stellar evolution models from the Modules for Experiments in Stellar Astrophysics code (MESA version 10398, \citealt{Paxton2011, Paxton2013, Paxton2015, Paxton2018}) that explore the subsequent long-term evolution of the disrupted star. The hydrodynamic simulations are explained in Section~\ref{subsec:hydro}, while the stellar evolution setup is detailed in Section~\ref{subsec:map}.

\begin{table}[ht]
    \caption{Summary of TDE remnant model properties}
 \begin{center}
    \begin{tabular}{c c c c c c}
    \hline
    \hline
        Name & Age & $\beta$ & $M_{i}$ & $M_{f}$ & Mass loss \\
        & (Gyr) & & ($M_{\odot}$) & ($M_{\odot}$) & (\%) \\
        \hline
        $1M\beta1$ & 4.8 & 1 & 1 & 0.956 & 4.4 \\
        $1M\beta1.5$ & 4.8 & 1.5 & 1 & 0.708 & 29.2 \\
        $1M\beta2$ & 4.8 & 2 & 1 & 0.386 & 61.4 \\
        $3M\beta4$ & 0.3 & 4 & 3 & 1.06 & 64.7 \\
        \hline
    \end{tabular}
    \end{center}
    \tablecomments{\footnotesize The TDE remnants are named for their initial mass and disruption parameter. Also listed are the stellar age at time of disruption in units of Gyr, disruption parameter $\beta$, progenitor mass $M_{i}$ in units of $M_{\odot}$, remnant mass $M_{f}$ in units of $M_{\odot}$, and percent mass lost due to the TDE. The disruption parameter is defined as $\beta \equiv \frac{R_{\tau}}{R_{p}}$ where $R_{\tau}$ is the tidal disruption radius and $R_{p}$ is the pericenter distance. High disruption parameters result in increased mass loss during the TDE.}
    \label{tab:models}
\end{table}

\subsection{Hydrodynamic FLASH models} \label{subsec:hydro}

In \cite{LawSmith2020}, stellar models of varying initial masses and ages from MESA were mapped into 3D FLASH simulations \citep{Guillochon2009, Guillochon2013}, where they were tidally disrupted by a SMBH with mass $M_{\rm BH} = 10^6\, M_{\odot}$.
The results of these simulations are cataloged in the STARS Library \citep{Zenodo2020}. For our study of the long-term stellar evolution post-disruption, we select three disruptions at solar metallicity ($Z = 0.0142$) of a star with initial mass $M_{i} = 1\, M_{\odot}$ which were tidally disrupted at an age of 4.8 Gyr. Based on the known population of stars in the Galactic Center \citep{Blum2003, Genzel2010, Pfuhl2011, Lu2013, Schodel2020}, we choose to primarily work with solar-mass stars to strike a middle ground between massive stars, which are rare, and low-mass stars, which are more abundant but are often fully disrupted, leaving no remnant. We selected middle-aged main sequence stars (MAMS) as our progenitors, as that is the evolutionary phase where stars spend the majority of their lives. We also model a fourth remnant from an initially $M_{i} = 3\, M_{\odot}$ progenitor, disrupted at an age of 0.3 Gyr (MAMS). The properties of each of the remnants modeled in this paper are summarized in Table \ref{tab:models}.

The models with a $M_{i} = 1\, M_{\odot}$ progenitor underwent partial tidal disruption with varying values of $\beta = 1$, 1.5, and 2. We define $\beta$ as the ratio of the tidal disruption radius ($R_{\tau}$) to the pericenter distance ($R_{p}$) -- a dimensionless parameter that characterizes the strength of the disruption. For instance, when $\beta = 1$ the pericenter distance is equal to the tidal radius, the point at which the SMBH's tidal forces first begin to siphon mass from the star. Above $\beta = 2$, the $M_{i} = 1\, M_{\odot}$ progenitors from the STARS Library are usually fully, or nearly fully, destroyed \citep[see Figure 5 in][]{LawSmith2020}. Thus for the $M_{i} = 1\, M_{\odot}$ progenitor, we select $\beta$ values distributed between 1 and 2 to best capture the possible range of disruption remnants. This set of remnant models represents the range of outcomes for common stars that are likely to be disrupted, while also being likely to survive. The corresponding final masses of each TDE remnant model from $M_{i} = 1\, M_{\odot}$ are  $M_{f} = 0.956, 0.708,$ and $0.386\,~M_\odot$ for $\beta=1$, 1.5, and 2 respectively. We refer to these models by their $\beta$ value and progenitor mass, e.g., the $1M\beta1$ remnant refers to the remnant disrupted at $\beta=1$ from a $M_{i} = 1\, M_{\odot}$ progenitor.

The $3M\beta4$ remnant is partially tidally disrupted with $\beta = 4$ from a progenitor with $M_{i} = 3\, M_{\odot}$. Post-disruption, the remnant mass is $M_{f} = 1.06\, M_{\odot}$, nearly the same as the initial mass of our other models. Its final mass is also very similar to the post-disruption mass of the $1M\beta1$ remnant. As a result, the long-term evolution of the $3M\beta4$ remnant provides an interesting contrast to that of similar-mass models with different initial conditions.

Figure \ref{fig:hydro} shows snapshots of the four TDE remnants at the end of the FLASH simulation, which terminates roughly 80 $t_{\rm dyn}$ after pericenter, where $t_{\rm dyn} \equiv ( G M_{\star}/R_{\star}^3)^{-1/2}$ is the star’s dynamical time \citep{LawSmith2020}. The three panels on the left compare the three remnants from the same progenitor of $M_{i} = 1\, M_{\odot}$. After the disruption, the least disrupted remnant ($1M\beta1$) retains a dense, hot core with a high internal diffusion time, and a relatively short diffusion time in the envelope. This suggests the TDE primarily deposited energy close to the star's surface without impacting much of the stellar interior. On the opposite end, the most disrupted remnant ($1M\beta2$) is relatively cooler throughout with a diminished diffusion time at the core, indicating the TDE had a greater impact on the star's internal structure. Energy deposited closer to the surface where the diffusion time is short will leak out first, powering the early-time luminosity of the remnant.

On the right, we show snapshots from the $3M\beta4$ remnant, whose radial extent is a factor of $\sim 2$ larger than the remnants from a $M_{i} = 1\, M_{\odot}$ progenitor. As this progenitor was disrupted at a very large $\beta$, the remnant also exhibits a short diffusion time throughout the core that signifies deep energy deposition from the TDE. Given the increased radial scale of these snapshots, we can see that the $3M\beta4$ remnant is much more diffuse, but still retains a relatively high core temperature due to its more massive progenitor. 

From the initial FLASH simulations, we map our chosen stellar remnants into MESA, using the method described in \citet{Wu2020}, and evolve them through time to investigate how the black hole encounter influences the remnants' evolution. From here onward, the four FLASH-to-MESA TDE remnant models described above will be referred to as ``disrupted stars'' or ``remnants.'' 

\begin{figure*}
\includegraphics[width=\textwidth]{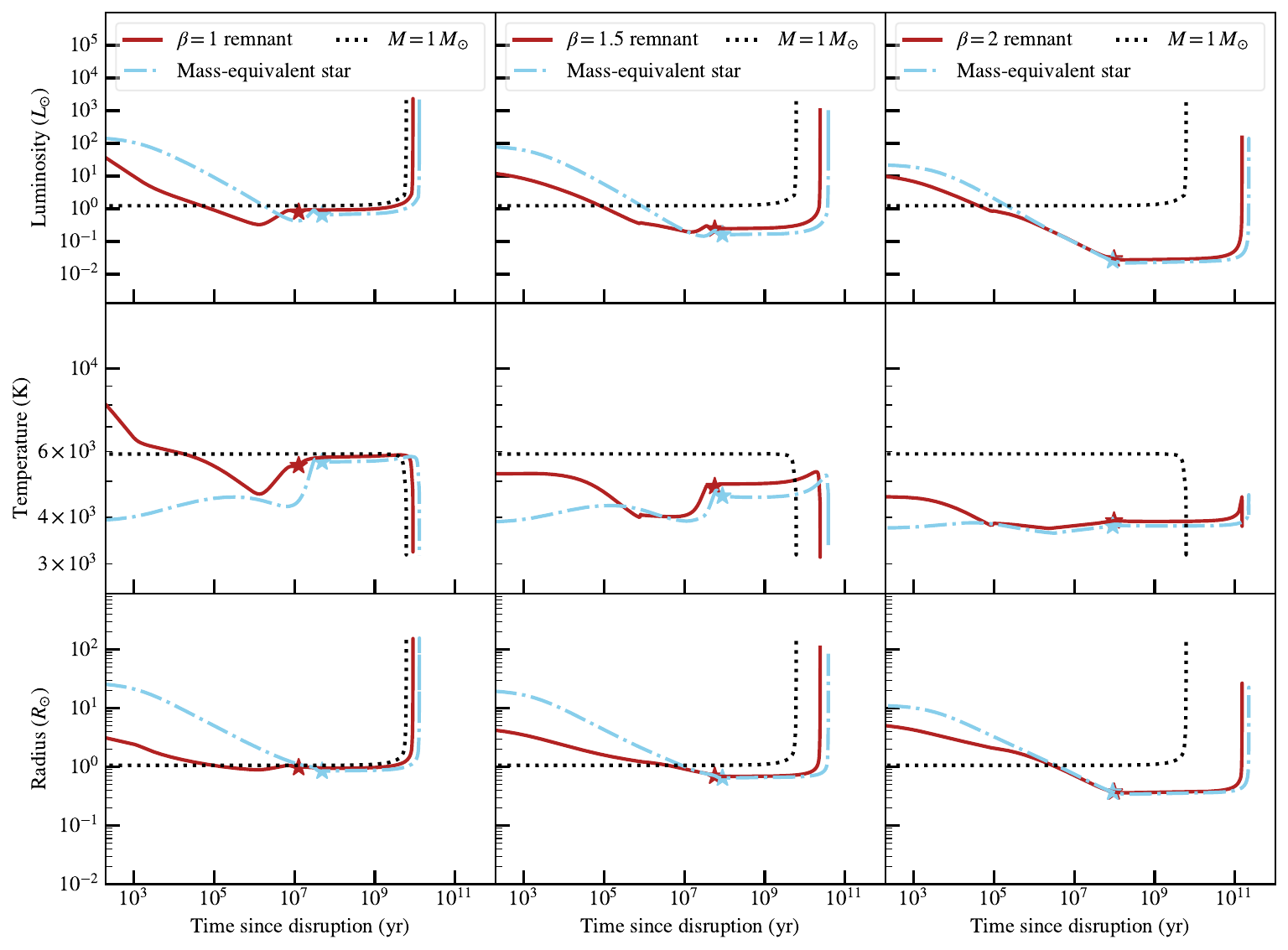}
\caption{Luminosity, temperature and radius evolution of the three tidally disrupted remnants (red solid) with progenitor mass $M_{i} = 1\, M_{\odot}$. These remnants are compared to mass-equivalent stars (blue dash-dotted) that evolve from the pre-main sequence (PMS) to the main sequence (MS) and up the red giant branch (RGB). The masses of the TDE remnants and their associated mass-equivalent stars are $0.956\, M_{\odot}$, $0.708\, M_{\odot}$, and $0.386\, M_{\odot}$ respectively from left to right. The start of core hydrogen fusion is denoted by star symbols for remnants and mass-equivalent stars. In addition, the evolution of an undisturbed $M = 1\, M_{\odot}$ star (black dotted) is shown from the time of disruption onward in order to represent the comparative evolution of the progenitor had it not undergone the TDE. Immediately after disruption, the remnants appear larger and more luminous than the progenitor, then shrink over the next several million years. After reigniting core hydrogen fusion, they stabilize near the MS locations of their mass-equivalent stars.
}
\label{fig:evol}
\end{figure*}

\subsection{Mapping into MESA} \label{subsec:map}

As described in Section~\ref{subsec:hydro}, the remnants include solar-mass stars disrupted at 4.8 Gyr with varying values of $\beta = 1$, 1.5, and 2, along with an initially $M_{i} = 3\, M_{\odot}$ star disrupted at 0.3 Gyr with $\beta = 4$ resulting in a $M_{f} \approx 1\, M_{\odot}$ remnant. For each remnant, we use the snapshot from the end of the FLASH simulation to map into MESA ($\approx80\, t_{\rm dyn}$ from pericenter, which corresponds to $\approx 38$ hr after pericenter for the $M_{i} = 1\, M_{\odot}$ progenitor and $\approx 124$ hr for the $M_{i} = 3\, M_{\odot}$ progenitor) \citep{LawSmith2020}. By this point in the FLASH simulation, the remnant's central density has relaxed to a steady state. We divide the domain of the final FLASH snapshot into material that is bound and unbound to the star; bound materials (including portions of the streams) become included as part of the remnant. We then use MESA's relaxation module to create an initial MESA model with a structure that closely matches the composition and entropy of the TDE remnant from the end of the FLASH simulation \citep[Appendix B,][]{Paxton2018}. We initiate our long-term stellar evolution simulations from this model. Our relaxation process allows us to observe the TDE remnant's further evolution from a much younger stage than is possible with other approaches to the mapping that begin following the remnant's evolution from the first few million years onward. We refer the reader to \citet{LawSmith2020} for further details on the stopping criteria in FLASH  and \citet{Wu2020} for the mapping procedure in MESA.

With the disrupted stars now mapped into MESA, we are able to run simulations of their long-term evolution. We adapt the simulation inlist from the MESA \texttt{1M\_prems\_to\_wd} test suite, which simulates the evolution of a 1 $M_\odot$, $Z = 0.02$ metallicity star from the pre-main sequence to a white dwarf. We adjust the metallicity to $Z = 0.0142$ \citep{Asplund2009}. Rotation is not included in this setup due to numerical limitations. We perform our analysis up until the turnoff to the white dwarf cooling sequence.

We are interested in comparing the evolution of TDE survivors with (1) their progenitors and (2) stars that are equal in mass to the remnants, which we call mass-equivalent stars. Consequently, we run a variety of MESA stellar evolutionary models of single, undisturbed stars in addition to the TDE remnant models listed in Table \ref{tab:models}. (1) To model the progenitors, we evolve undisturbed stars with zero age main sequence (ZAMS) masses equivalent to the initial masses of the remnants before the TDE occurred (i.e., $M = 1\, M_{\odot}$ and $3\, M_{\odot}$). (2) To model mass-equivalent stars, we  evolve undisturbed stars with ZAMS masses equivalent to the final masses of the remnants after the TDE (i.e.,  $M = 1, 0.956, 0.708, {\rm and\,} 0.386\, M_{\odot}$).  
Our complete MESA inlists with all of our parameter adjustments can be found on \texttt{Zenodo} \citep{Zenodo2025}.

\section{Results} \label{sec:results}

Each remnant loses a portion of its envelope to the SMBH during the disruption, ranging from just a few percent to over two thirds mass loss, depending on the disruption parameter $\beta$ (see Table \ref{tab:models}). 
In Section~\ref{subsec:early}, we will first examine the early rapid evolution that occurs during the initial bright phase of the TDE remnants; in Section~\ref{subsec:nitrogen}, we detail the effects of the TDE on remnants' colors and surface abundances; and in Section~\ref{subsec:long}, we describe the stable behavior that the TDE remnants settle into over long timescales.

\subsection{Early Stellar Evolution} \label{subsec:early} 

Previous studies \citep{Guillochon2013,LawSmith2020} demonstrate that the work done by tidal forces injects energy into the layers of the remnant. Various slices of the star are torqued by the differential gravitational pull of the SMBH, imparting angular momentum and causing shearing that heats the star's interior. The SMBH also pulls stellar material out to larger radii, causing it to be less bound. We expect from hydrodynamical simulations that the infalling mass accretion imparts additional energy in and around the surviving core \citep{LawSmith2020}. This extra energy is retained in the remnant that we map into MESA.

Zooming in on the early behavior for the remnants resulting from disruption of a $M_{i}= 1\, M_{\odot}$ progenitor, Figure \ref{fig:evol} shows that rapid evolution occurs in the remnants' luminosities, effective temperatures, and radii during the first $10^3$ yr after disruption. The injection of energy from the tidal encounter causes the luminosities of the remnants (solid red lines) to soar orders of magnitude higher than that of their respective progenitors (black dotted): the luminosity of the $1M\beta1$ remnant increases by a factor of $\approx 500$, and the luminosities of the $1M\beta1.5$ and $1M\beta2$ remnants increase by a factor of $\approx 10$. 

We note that the $1M\beta1$ remnant experiences the highest jump in luminosity, which we attribute to the fact that its radius increases greatly while retaining almost all of its original mass. The $1M\beta1$ remnant also experiences the steepest decline in luminosity, which relates to where the energy was deposited and how long it takes to be released. The instantaneous luminosity of the remnant will be $\approx$ (energy deposited)/(diffusion time) for the region that is producing the luminosity, with the outer layers with short diffusion times producing the radiation first. Even though the $\beta = 2$ encounter deposits more energy overall, the initial luminosity of the $1M\beta2$ remnant is still lower than the initial luminosity of the $1M\beta1$ remnant, which implies that less energy is deposited in the envelope where the diffusion time is short for $\beta = 2$ than for $\beta = 1$. In mild TDEs where the star just grazes the SMBH's tidal radius (as for $1M\beta1$), most of the energy is deposited in the outer envelope where it is quick to diffuse out, whereas more severe disruption events (as for $1M\beta1.5$ and $1M\beta2$) result in energy deposited deeper in the interior where it takes longer to leak out. This may explain why the $1M\beta1$ encounter is initially the brightest, but also fades most rapidly.

The remnants also grow $4$--$6$ times in radius compared to the progenitor stars. The inflated envelopes they develop during the TDE can also be seen in the FLASH snapshots in Figure \ref{fig:hydro}. Directly after the TDE, the $1M\beta1.5$ and $1M\beta2$ remnants exhibit surface temperatures lower than that of their progenitor as a result of their diminished mass and puffy envelope (top panels of Figure \ref{fig:hydro}, columns 2-4). The $1M\beta1$ remnant does not follow this trend (top panel of Figure \ref{fig:hydro}, column 1), having experienced a less severe TDE leading to relatively low mass loss, and instead shows a highly elevated temperature.

We can also compare the early evolution of the TDE remnants to each remnant's mass-equivalent star (blue dash-dot). We evolve the mass-equivalent stars as they contract along the pre-main sequence (PMS), move onto the main sequence (MS), then finally ascend the red giant branch (RGB). To facilitate comparison in Figure~\ref{fig:evol}, we have shown the mass-equivalent stars as beginning their PMS at $t = 0$, which represents the time at which the remnants were disrupted. All three remnants of the $M_{i} = 1\, M_{\odot}$ progenitor appear initially hotter than the mass-equivalent stars on the PMS, before cooling off to similar temperatures as the mass-equivalent stars on the MS. Once hydrogen fusion is ignited in either the remnants or the mass-equivalent stars, they proceed on the main sequence and then up the RGB in quite similar ways, although the TDE remnants tend to remain slightly brighter and hotter than their mass-equivalent stars.

During the first $10^5$ yr, these low-mass remnants appear to be masquerading as younger, more massive stars due to their high temperatures and luminosities. This is the stage when the disrupted stars look most distinctive from their undisturbed progenitors, and although exciting, it does not last. This early appearance fades within roughly Kelvin-Helmholtz timescales as the excess energy leaks out of the remnant and it returns to the luminosity level of its progenitor (see Appendix~\ref{app:early_evolution} for additional details on the early post-disruption evolution). Of the estimated one million tidally disrupted remnants we expect the Milky Way to harbor (see Equation \ref{equation1}), the number currently traversing this brief, bright stage should be on the order of 10: for a partial TDE rate of $\approx 10^{-4}\, \mathrm{yr}^{-1}$, we estimate that $\lessapprox10$ remnants can be observed in their first $10^5$ yr post-disruption (see Equation \ref{equation2}). 
\begin{equation} \label{equation1}
    10^{-4}\, \mathrm{remnants /yr}\times 10^{10}\; \mathrm{yr} = 10^6\; \textrm{remnants}
\end{equation}

\begin{equation} \label{equation2}
    10^{5}\, \mathrm{yr} \times 10^{-4}\, \mathrm{remnants/yr} = 10\; \textrm{remnants}
\end{equation}

Following this initial bright period, the remnants continue contracting for $\approx 10^7$--$10^8$ yr, consistent with the core's thermal timescale, until they become dense enough to re-ignite hydrogen core fusion, ultimately converging to mimic the MESA models of their lower mass-equivalent stars. The vast majority of TDE survivors today should have already settled into their late-stage appearance, as we discuss in Section~\ref{subsec:long}.

\begin{figure}
\includegraphics[width=\columnwidth]{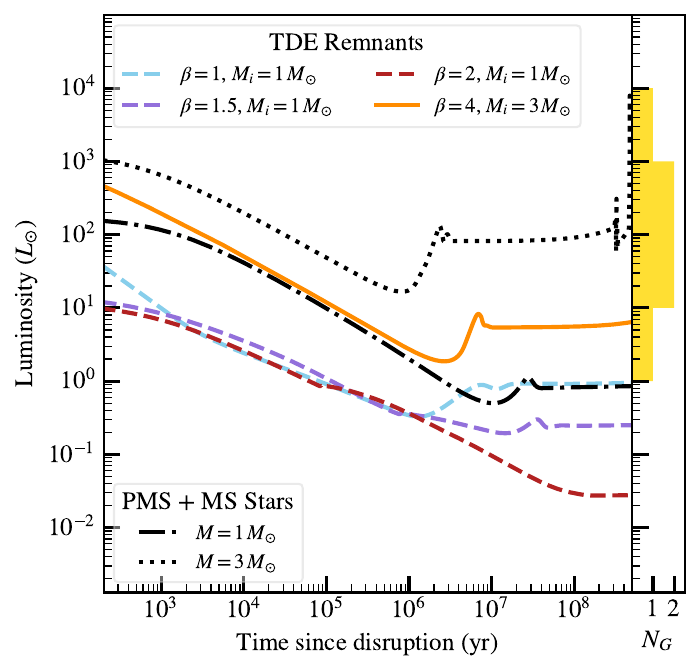}
\caption{Luminosity evolution for the first few 100 Myr of the TDE remnants. The $1M\beta1$, $1M\beta1.5$, and $1M\beta2$ TDE remnants from Figure \ref{fig:evol} are shown with dashed lines and are compared here with their $1\, M_{\odot}$ progenitor (black dash-dotted). Additionally, the $3M\beta4$ remnant is shown (solid) to remain brighter than its mass-equivalent star (black dash-dotted), but dimmer than its progenitor (black dotted), throughout the PMS + MS. On the right, a histogram of the number of G-type objects ($N_G$) per luminosity bin is shown (data from \citealt{Ghez2004, Ghez2005, Hornstein2007, Phifer2013, Sitarski2016PhDT, Witzel2014}).}
\label{fig:Levol_allmodels}
\end{figure}

Figure \ref{fig:Levol_allmodels} compares the early luminosity evolution of each TDE remnant from the $1\, M_{\odot}$ progenitor to that of the remnant from the $3\, M_{\odot}$ progenitor. The evolution of undisturbed stars on the PMS and early MS with $M = 1\, M_{\odot}$ and $3\, M_{\odot}$ are also shown for comparison; these represent the progenitors for our four TDE remnants, and the $M = 1\, M_{\odot}$ model also serves as a mass-equivalent star for the $3M\beta4$ remnant. The $3M\beta4$ remnant is less luminous throughout its post-disruption lifetime relative to its progenitor, which can to a large extent be attributed to its 64.7\% reduction in mass. Although it has roughly the same final mass as a $1\, M_{\odot}$ star, the remnant's luminosity is initially larger than its mass-equivalent star by a factor of a few. Its luminosity decreases during the early ages of $\approx\, 10^7$ yr until it ignites hydrogen fusion, then its luminosity stabilizes to be $\approx 5$ times greater than that of the mass-equivalent star on the MS. 

The behavior of the $3M\beta4$ remnant can be contrasted with the behavior of the remnants disrupted from the $M_{i} = 1\, M_{\odot}$ progenitor. The $1M\beta1$, $1M\beta1.5$, and $1M\beta2$ remnants return to much more similar luminosities, temperatures, and radii once hydrogen is ignited when compared with their mass-equivalent stars (see Figure \ref{fig:evol}). In contrast, the $3M\beta4$ remnant retains a longer-lasting signature of its history via a significantly enhanced luminosity during hydrogen burning. Though the $3M\beta4$ remnant has a final mass of $M_{f} = 1.06\, M_{\odot}$ that is very similar to the final mass $M_{f} = 0.956\, M_{\odot}$ of the $1M\beta1$ remnant, the latter experienced a less disruptive encounter and does not show a similarly enduring imprint of the TDE during the hydrogen burning phase.

As the results of encounters that are common in the Galactic Center, the TDE remnants we model here are potential candidates for the unusual G objects observed in the nuclear star cluster of the Milky Way \citep{Gillessen2012, Witzel2014, Ciurlo2020}. The right panel of Figure \ref{fig:Levol_allmodels} depicts a histogram of the number of G objects per luminosity bin \citep[using data from][]{Ghez2004, Ghez2005, Hornstein2007, Phifer2013, Sitarski2016PhDT, Witzel2014}. The luminosities of the brightest G objects are not attained by any of our remnants, though the $3M\beta4$ remnant does achieve a luminosity comparable to the $10^2$--$10^3\, L_\odot$ bin during the first few thousand years after disruption. The lower-$\beta$ remnants from $M_{i} = 1\, M_{\odot}$, which we expect to be more common, match the luminosity of the faintest G objects ($1$--$10^2\, L_{\odot}$) during the first $\approx 10^5$ yr after disruption. Although not all G objects fall within our TDE remnant luminosity ranges, we are interested to see that some do.

\begin{figure}
\includegraphics[width=\columnwidth]{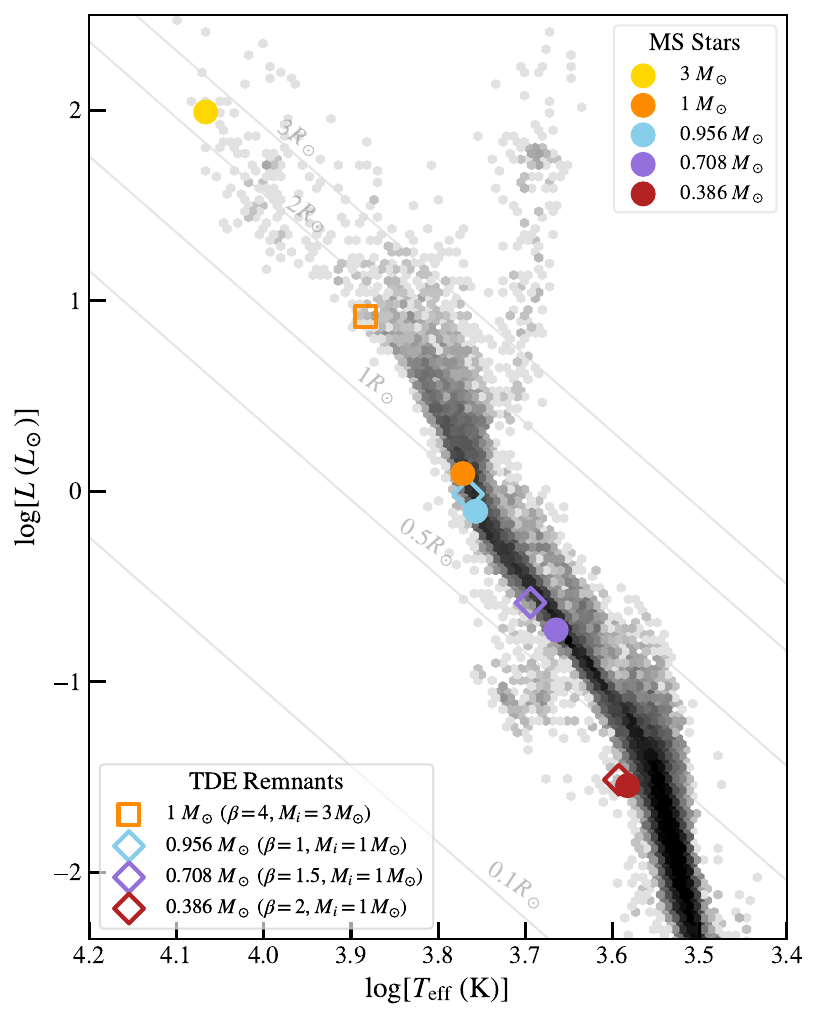}
\caption{Hertzsprung-Russell diagram of simulated stars on the hydrogen-burning main sequence. Data from \textit{Gaia DR3} \citep{Gaia2016, Gaia2021} of the closest 50,000 stars is plotted in the background as a logarithmic histogram where grayscale shading indicates number density, to show the natural spread of Milky Way stars. The filled circles are main sequence single stars simulated in MESA, with the ZAMS masses listed in the upper right legend. The open diamonds represent the $1M\beta1$, $1M\beta1.5$, and $1M\beta2$ TDE remnants, and the open square is the $3M\beta4$ TDE remnant; all are shown at snapshots when they have returned to stable hydrogen burning, and the final masses of each remnant are noted in the bottom legend. All remnants appear substantially cooler and dimmer than their respective progenitors, while appearing subtly hotter and brighter than their mass-equivalent stars (or extremely hotter and brighter, in the case of the $3M\beta4$ remnant).
\label{fig:hr}}
\end{figure}

\begin{figure*}
\includegraphics[width=\textwidth]{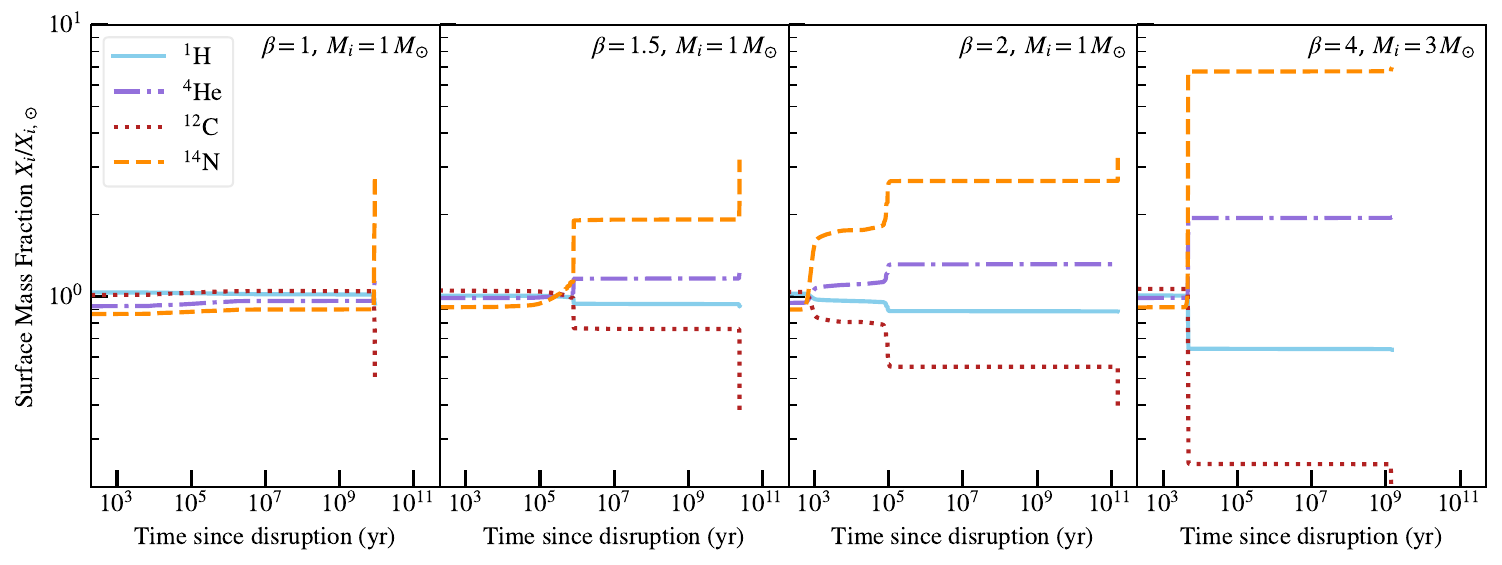}
\caption{Surface mass fractions of selected elements $X_{i}$ in our remnants, scaled to solar metallicity abundances $X_{i,\odot}$. We show results for the elements $X_{i}$ listed in the legend. Lines end just before core He ignition. The $1M\beta1$ remnant retains near-solar abundances until a dredge-up phase when it becomes a red giant. For the disrupted stars with $\beta > 1$, by the main sequence the envelopes are enriched in nitrogen (orange dashes) and depleted in carbon (red dots), both indicators of CNO fusion processes in the core. The presence of core-like CNO ratios on the stars' surfaces is a result of convective mixing that occurs during the first $10^5$ yr after disruption. Enhancement of helium is also observed in each of these models. These differences may manifest as spectroscopic signatures in the remnants of partial TDEs.
\label{fig:abund}}
\end{figure*}

\subsection{Colors and Abundances} \label{subsec:nitrogen}

The TDE remnants eventually come to settle into a stable period of hydrogen fusion which lasts for $10^9$--$10^{11}$ yr, which can be interpreted as them returning to the behavior of stars on the main sequence. When placed on a Hertzsprung-Russell diagram (Figure~\ref{fig:hr}), survivors are offset from what would be expected. Our models  appear cooler and less luminous (redder) than their corresponding progenitor, but also hotter and more luminous (bluer) than mass-equivalent stars. For TDE remnants originating from a $M_{i} = 1\, M_{\odot}$ progenitor, the offset of the TDE remnants from their mass-equivalent stars is subtle. However, the offset is more extreme for the $3M\beta4$ remnant, which is an order of magnitude brighter and $> 1000$ K hotter than its $1\, M_{\odot}$ mass-equivalent star. This results from the remnant possessing a denser and hotter core, such that it fuses hydrogen via both pp-chain and CNO fusion in almost equal measure. However, all remnants fall within the natural deviation of stars on the main sequence observed by \textit{Gaia}; therefore, without an independent mass measurement, color alone is not a conclusive method for finding TDE survivors.

An independent avenue for identifying survivor stars would be searching for evidence of atmospheric enrichment. Simulations show that TDEs can effectively mix the composition of the star, dislodging helium and heavier elements such as carbon and nitrogen from the core \citep{Kochanek2016, Gallegos-Garcia2018, LawSmith2019, Mockler2022}. Subsequent to the TDE, each remnant develops convection in the envelope as it evolves, transporting some of these heavier elements to the surface. Figure \ref{fig:abund} shows surface abundances scaled to solar up until just before core He ignition. Focusing first on the stars disrupted from a $1\, M_{\odot}$ progenitor, the least disturbed remnant ($1M\beta1$) exhibits convection that is confined to the envelope. Its abundances stay roughly equivalent to solar, until it experiences dredge-up as it enters the red giant branch. However, the $1M\beta1.5$ and $1M\beta2$ remnants develop progressively larger convective zones in the $10^5$ years after disruption. Figure \ref{fig:hydro} demonstrates that the material of the $1M\beta2$ remnant in FLASH is cooler throughout than that of the lower-$\beta$ remnants. This translates into cooler temperatures for the MESA model of the remnant, and leads to the formation of a deeper convective zone. Such an effect is in line with expectations that the depth of the convective envelope increases with decreasing ZAMS mass for undisturbed stars of the same masses as the remnants. The presence of convection in these remnants promotes further mixing of elements from the bottom of the convective zone, which include products from the CNO burning which occurred in the core of the $1\, M_{\odot}$ progenitor before disruption. Thus, the degree of element enrichment increases with the $\beta$-parameter for a given initial progenitor mass, because stars that approach closer to the SMBH experience more severe disruption that produces diffuse, cooler remnants with deeper convection. 

We find that this results in an envelope abnormally enriched in helium for remnants with deep convective envelopes (Figure \ref{fig:abund}) as compared to the surface helium abundance expected of typical single stars at solar metallicity. After disruption and once the remnant has reignited core hydrogen-burning, we find that the $1M\beta1.5$ and $1M\beta2$ remnants have enhanced He abundance relative to solar (using the \citealt{Asplund2009} solar abundances) ranging from $X_{\rm He}/X_{\mathrm{He},\odot}\approx 1.2$--$1.3$.
Similarly, the envelope may become enriched in nitrogen and depleted in carbon due to mixing. Figure \ref{fig:abund} shows that the difference in carbon and nitrogen mass fractions compared to a Sun-like star can be significant, with $X_{\rm N}/X_{\mathrm{N},\odot} \approx 3.2$ and $X_{\rm C}/X_{\mathrm{C},\odot} \approx 0.4$. This difference in envelope composition will likely affect the appearance of stellar spectra and potentially provides a secondary diagnostic tool to find these remnants. The mixing seen in MESA which manifests as each jump in the abundance in Figure 5 is sensitive to choices of convection parameters and other numerical considerations. It takes approximately $10^5$ yr for the mixing to conclude, after which the surface ratios remain stable until the end of the main sequence. The exact time evolution of these abundance ratio values before the mixing concludes should be taken as approximate.

The altered composition of the remnant is especially noticeable for the $3M\beta4$ remnant, shown in the right-most panel of Figure \ref{fig:abund}. Compared to a solar-mass star at solar abundance, which has a similar total mass to this remnant, each of the hydrogen, helium, carbon, and nitrogen abundances at the surface are found to be appreciably altered ($X_{\rm He}/X_{\mathrm{He},\odot} \approx 2$, $X_{\rm N}/X_{\mathrm{N},\odot} \approx 7$, $X_{\rm C}/X_{\mathrm{C},\odot} \approx 0.2$). This is still caused by convective mixing after the TDE, but the effect is enhanced since the $3\, M_{\odot}$ progenitor primarily burned hydrogen on the main sequence via the CNO cycle before the TDE occurred. For these highly disrupted, higher-mass main sequence stars, we expect an even greater prospect of distinguishing a history of tidal disruption through an unusual stellar spectrum. Our findings are corroborated by a similar study completed by \citet{Sharma2024}, in which they simulate the evolution of TDE remnants of $M_{i} = 1$ and $M_{i} = 3\, M_{\odot}$ MAMS stars and also find that high-$\beta$ remnants exhibit compositional mixing that leads to heightened surface enrichment of N and He (Figures 13, 14, and 18 in \citealt{Sharma2024}).

\subsection{Long-Term Stellar Evolution} \label{subsec:long}

Our three remnants disrupted from a $1\, M_{\odot}$ progenitor lose varying degrees of mass during the tidal disruption event and consequently spend the rest of their lives burning less intensely. This lowered rate of fusion resulting from their mass loss allows them to live billions of years longer than their progenitor. While the $1\, M_{\odot}$ progenitor ages off of the main sequence around age 11 Gyr (roughly 6 Gyr after disruption), the $1M\beta1$ remnant's life is extended by an additional 3 Gyr; by 19 Gyr for the $1M\beta1.5$ remnant; and by 140 Gyr for the $1M\beta2$ remnant, as seen in Figure \ref{fig:evol}. Conversely, each remnant has a shorter lifetime than its mass-equivalent star, which is consistent with the remnant having the core density of a higher-mass star. 

Once the $3M\beta4$ remnant reignites core hydrogen fusion, it similarly lives longer than its progenitor and reaches the RGB earlier than its mass-equivalent star. Notably, the remnant's luminosity during hydrogen burning is larger than that of its mass-equivalent star by a factor of $\approx 10$ during its long-term evolution on the main sequence, which is a much larger contrast than any of the remnants disrupted from a $M_{i} = 1\, M_{\odot}$ progenitor achieve. Consequently, it may be possible to differentiate the TDE remnant of a heavily disrupted high-mass star (e.g., $3M\beta4$) from the remnant of a mildly disrupted low-mass star (e.g., $1M\beta1$), in that a more heavily disrupted high-mass remnant maintains a heightened luminosity signature for its entire life, even after having reignited core hydrogen fusion.

While our work is most concerned with the previously unstudied early-time behavior of the remnants, we evolve our models self-consistently until the end of their time on the main sequence to offer a holistic picture of their entire evolution from initial conditions to old age. We find that our early-time behavior leads naturally to late-time behavior that is consistent with \citet{Sharma2024}, which studied the late-time evolution in detail. At late times, both of our studies find that the remnant is cooler and dimmer than its progenitor, yet hotter and brighter than its mass equivalent star; that the remnants resulting from higher-$\beta$ interactions live longer; and that the brightest remnants come from low-$\beta$ encounters that lose the least amount of mass.

We note that our study differs from \citet{Sharma2024} in that our hydrodynamical FLASH calculations have 10$\times$ higher resolution. This allows us to effectively capture the early evolutionary behavior for $t\lesssim 10^5$ yr, during which time the TDE remnant is brightest and may be most visible. There is value in modeling the early evolution of the remnant, as it provides a sense of how the mixing evolves over time, showcases their brightest and hottest phase, and opens the question of whether certain rare objects and bright transients might be explainable as young TDE remnants.

\section{Discussion and Conclusions} \label{sec:disc}

In this work, we investigate the post-disruption evolution of partial TDE remnants in order to develop strategies for discovering these objects. Our models indicate the remnant will be brightest in the first $10^5$ yr after disruption, during which period they may appear unusually luminous and extended compared to undisturbed stars. Over timescales of $\approx 10^7$--$10^8$ yr, however, the majority of TDE remnants will have long faded to appear quite similar to main sequence stars of a similar mass. The (relatively uncommon) high-$\beta$ remnant from a more massive progenitor does appear substantially offset in luminosity and temperature from its mass-equivalent, but all remnants are still expected to fall within the natural variation of MS stars on a Hertzsprung-Russell diagram. Nevertheless, TDE remnants constitute a class of interesting objects which may be identified observationally in a few key ways.

\subsection{Early Elevated Luminosity}
\label{subsec:discuss_lum}

We are particularly interested in similarities between our models of TDE remnants and the known class of G objects in the Galactic Center. The G objects move with star-like kinematics, but have massive dust-shrouds that make them appear more like clouds. Although they are yet to be explained, the G objects pose interesting candidates for young TDE survivors, given their location in the Galactic Center, elevated luminosities, and puffy radii. The bolometric luminosities of G objects are commensurate with those of the brightest TDE remnants. Figure~\ref{fig:Levol_allmodels} compares the inferred blackbody (bolometric) luminosities of the six G objects reported in \cite{Sitarski2016PhDT} with the early luminosity evolution of our TDE remnant models. Although their luminosity distributions are not coincident, there is some overlap between the dimmest G objects, the higher-mass TDE remnant at any age, and the lower-mass TDE remnants during their first $10^5$ yr post-disruption.

In order to understand whether TDE remnants may be related to G objects, it is important to estimate whether TDEs occur at a rate high enough to suggest that young, high-luminosity remnants could account for some or all of the current population of G objects. Partial TDEs are expected to occur at rates up to $10\times$ higher than full disruptions; for instance, in the full loss cone regime, the rate of TDEs should scale with the inverse of $\beta$, such that low-$\beta$ encounters resulting in partial disruptions can be expected to occur much more frequently than high-$\beta$ encounters resulting in full disruptions \citep{Krolik2020, Stone2016, Stone2020, Bortolas2023}. However, the heightened partial TDE occurrence rate of $\sim 10^{-4}\, \mathrm{yr}^{-1}$ is countered by the fact that our models show partial TDE remnants fading rapidly within $\lesssim 10^5$ yr. Accounting for both effects, we estimate that at any given moment in the Galactic Center, we can expect a handful of TDE remnants to exist that have high enough luminosities to match those of the dimmest G objects. Thus, we predict that TDE remnants in their early phase of evolution could account for a few G objects, but not the entire population \citep{Sitarski2016PhDT}.

To further explore the potential connection between TDE remnants and G objects, it would be worthwhile for future studies to incorporate the effect of dust. G objects have been observed to have temperatures much cooler than would be expected for their luminosities, suggesting the light of an enclosed star may be getting reprocessed to longer wavelengths by a surrounding cloud of gas and dust \citep[e.g.,][]{Witzel2014}. Our hydrodynamic FLASH simulation did not incorporate the effects of dust, so we cannot speak to the impact of TDEs on the production and distribution of dust in remnants, nor the dust's subsequent evolution, but this would be a fruitful avenue for future exploration.

\subsection{Kinematics \& Spectroscopy} \label{subsec:discuss_kine_spec}

Given that the vast majority of TDE remnants have aged past their initial energized state, what are the best criteria to use when attempting to identify an older remnant after it has faded? We recommend employing a combination of kinematics and spectroscopy. 

There exist several possible kinematic outcomes for TDE remnants. Disrupted stars on elliptical orbits around the SMBH will remain bound as remnants \citep{Manukian2013} and may return periodically to experience repeated disruptions by the SMBH \citep{MacLeod2013, Cufari2022, Liu2023, Bandopadhyay2024, Liu2025}. However, those remnants that maintain parabolic or hyperbolic orbits will eventually exit the sphere of influence of the SMBH \citep{Manukian2013, Ryu2020, Sharma2024}, traveling radially outward. Interactions between the star and the SMBH can lead to exchanges in mass and momentum, which for certain geometries may result in ``turbovelocity'' stars \citep{Manukian2013}, suggesting that many TDE remnants may pass through the Galactic Center at speeds of a few $\times 100$ km s$^{-1}$. Stars are even found in the Galactic halo moving at elevated speeds $> 1000$ km s$^{-1}$, known as ``hypervelocity'' stars. Some hypervelocity stars may originate from the disruption of binary systems by the SMBH in the Galactic Center \citep{Hills1988,Manukian2013,Brown2015, Brown2018}. However, recent work analyzing the spatial and kinematic distributions of hypervelocity stars has shown that they may have an alternate origin, tracing back to the Large Magellanic Cloud instead of the Galactic Center \citep{Generozov2022,Han2025arXiv}.

What are the likely kinematics of our remnants? The stars in this study arrive on parabolic orbits according to the \cite{LawSmith2020} FLASH simulation, marginally bound to the SMBH, and receive a kick during the disruption. Previous work from \cite{Manukian2013} has demonstrated that the remnants of partial disruptions of single stars arriving on parabolic orbits will likely remain bound to the Milky Way, as they are not able to receive large enough kicks to result in ejection of the star from the Galaxy. \cite{Manukian2013} predict a population of stars in the Galactic Center with velocities that are similar to turbovelocity stars, also consistent with \cite{Ryu2020}. From \cite{Manukian2013}, we expect our remnants could be contributing to the population of turbovelocity stars, but likely not the hypervelocity stars, as our remnants are not unbound from the Galaxy. Our remnants will likely remain on similar orbits to their original approach, and the kick they receive due to the TDE is usually large enough that the remnants avoid a second disruption.

Thus, given a selection of fast-moving stars with trajectories emanating from the Galactic Center, we would then suggest a spectroscopic follow up study could be conducted to look for He or N enrichment in the stellar envelope. Such enrichment indicates that the candidates have experienced an extreme mixing event like a TDE. In the future, one fruitful avenue would be to simulate the expected spectra of TDE remnants to identify specific spectral fingerprints. The characteristic He and N surface enrichment of remnants will persist long after the remnants have faded from their initial luminous phase, therefore spectral signatures could be an especially useful diagnostic with the older remnant population.

\subsection{Summary}
\label{subsec:discuss_sum}

In summary, we believe that searching for TDE survivors in the Milky Way is a compelling opportunity to probe gravitationally intense interactions from a new angle. The key findings of this work are the following:\\

\begin{itemize}
    \item The survivor stars are initially highly altered compared to their progenitors, losing mass at a rate determined by the strength of the encounter, and increasing in radius, temperature, and luminosity, albeit for a brief period of time ($\lesssim 10^5$ yr). The likelihood of finding these remnants during their bright phase is low; we estimate, to within an order of magnitude, that there are roughly 1--10 such remnants in the Milky Way. During this phase, survivors have properties similar to those observed for G objects in the Galactic Center. 
    
    \item The remnants cool on a timescale of $10^7$--$10^8$ yr until they reignite hydrogen fusion in their cores and reach an equilibrium state consistent with the main sequence, and are therefore difficult to differentiate by color alone. The vast majority of survivors exist in this equilibrium phase due to its longevity. 

    \item More heavily disrupted high-mass remnants maintain a heightened luminosity signature. Thus it may be possible to differentiate the TDE remnant of a heavily disrupted high-mass star from the remnant of a mildly disrupted low-mass star that both result in the same final mass. 
    
    \item Due to the intense mixing that occurs during and after the TDE, helium- and nitrogen-enriched envelopes result in all except the least-disrupted remnant. As such, spectroscopic characterization might be the most effective way of uncovering the survivor population in the Galactic Center.
\end{itemize}

The introduction of the Vera C. Rubin Observatory Legacy Survey of Space and Time, coming online in the next year, provides an exciting avenue for possibly observing even more TDEs, including partial disruptions, among the expected flood of newly captured transient data \citep{Bricman2020}. To complement the influx of new TDEs, it is important to understand what remains after the transient dims. Searching for TDE remnants requires a different strategy from capturing the bright and brief flash of a TDE itself, but we hope that studies like our own \citep[e.g.,][]{Sharma2024} that model the detailed evolution of a TDE remnant may illuminate critical characteristics that will help with the identification of TDE remnants among the stars.

\begin{acknowledgments}
We thank Andrea Ghez, Charles Gibson, Ylva G{\"o}tberg, Erika Holmbeck, Fulya Kiroglu, Jamie Law-Smith, James Lombardi, Morgan MacLeod, Brenna Mockler, Ruth Murray-Clay, Smadar Naoz, Fred Rasio, Sanaea Rose, and Sarah Wellons for many enlightening discussions. The authors thank Victoria Molero Gonz\'alez, Deana Tanguay, and all those involved with the Lamat Institute who invest their energy into co-creating a supportive scientific community. This research was made possible thanks to funding from the Lamat Institute and UC Santa Cruz through the Heising-Simons Foundation, NSF grants: AST 1852393, AST 2150255 and AST 2206243. S.C.W. is grateful for support from the National Science Foundation Graduate Research Fellowship under Grant No. DGE‐1745301. R.W.E. gratefully acknowledges the support of the Heising-Simons Foundation, the Vera Rubin Presidential Chair for Diversity at UCSC, and NASA ATP Grant 80NSSC22K0722 to UNC at Chapel Hill. R.Y. acknowledges support from a NASA FINESST award (21-ASTRO21-0068).  A.M-B.\ is supported by NASA through the Hubble Fellowship grant HST-HF2-51487.001-A awarded by the STScI, which is operated by the AURA under NASA contract NAS5-26555. We acknowledge use of the lux supercomputer at UCSC, funded by NSF MRI grant AST 1828315. We acknowledge use of the computational resources of the SCIENCE HPC Center at the University of Copenhagen, with the support of DARK at NBI. The 3D hydrodynamics software used in this work was developed in part by the DOE NNSA- and DOE Office of Science-supported Flash Center for Computational Science. This work has made use of data from the European Space Agency (ESA) mission {\it Gaia} (\url{https://www.cosmos.esa.int/gaia}), processed by the {\it Gaia}
DPAC (\url{https://www.cosmos.esa.int/web/gaia/dpac/consortium}). Funding for the DPAC has been provided by national institutions, in particular the institutions participating in the {\it Gaia} Multilateral Agreement. We acknowledge that the land on which UCSC is located, and where this research was conducted, is the unceded territory of the Awaswas-speaking Uypi Tribe. The Amah Mutsun Tribal Band, comprised of the descendants of Indigenous people taken to missions Santa Cruz and San Juan Bautista during Spanish colonization of the Central Coast, is today working hard to restore traditional stewardship practices on these lands and heal from historical trauma. We affirm that through our research and relationships, we have an obligation to deepen the wellness of our communities–those we come from; those we share this planet with now, including those impacted by our work, telescopes, and facilities; and those coming after us. 
\end{acknowledgments}

\software{\texttt{Astropy} \citep{Astropy2013, Astropy2018}, \texttt{FLASH} \citep{Fryxell2000}, \texttt{Matplotlib} \citep{Hunter2007}, \texttt{MESA} \citep{Paxton2011, Paxton2013, Paxton2015, Paxton2018}, \texttt{Numpy} \citep{Numpy2020}, \texttt{STARS\_library} \citep{Zenodo2020}.}

\appendix 
\vspace{-0.5cm}

\section{Early evolution of weak encounters: A contracting envelope model} \label{app:early_evolution}

\begin{figure*}[b]
    \centering
    \includegraphics[width=0.50\textwidth]{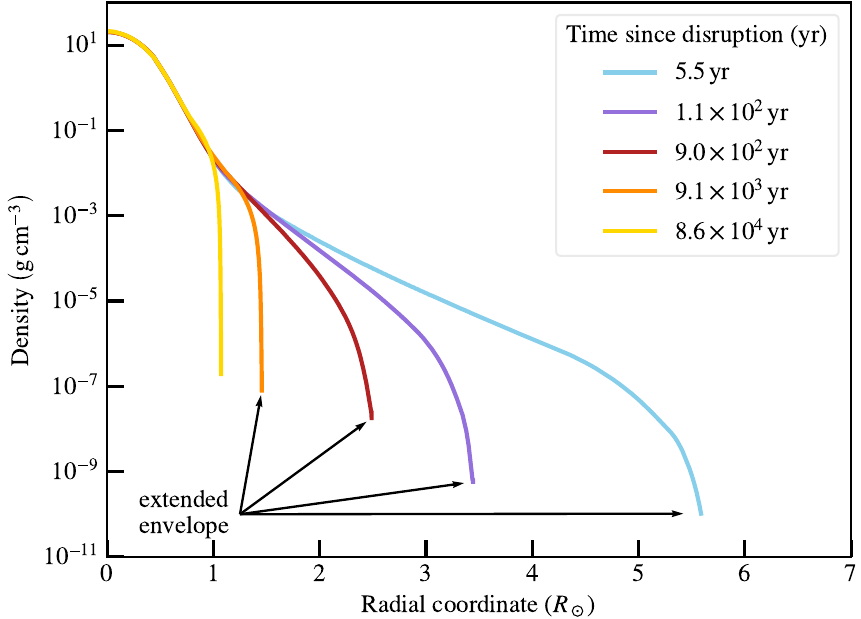}
    \includegraphics[width=0.475\textwidth]{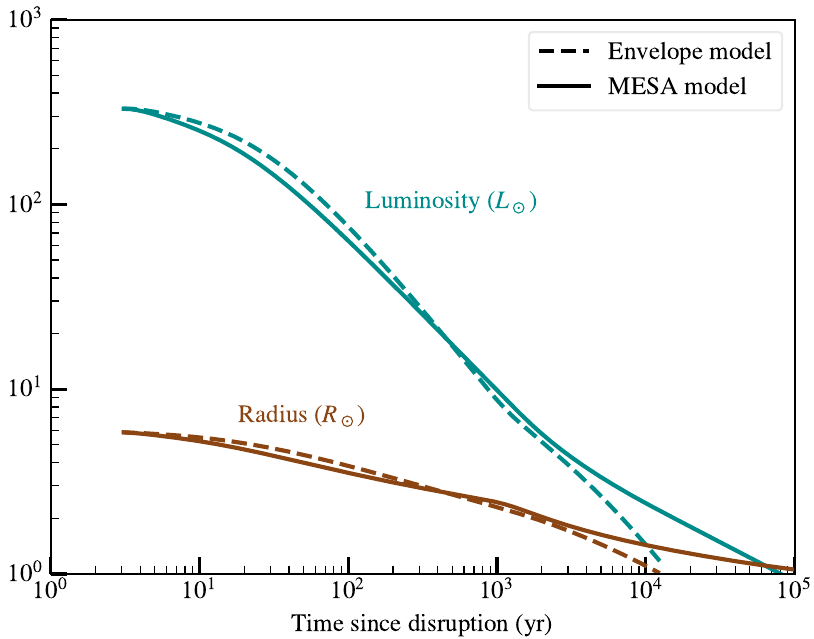}
    \caption{Left panel: Density profile of the remnant at several points in time after a $\beta = 1$ encounter. During the encounter, the star becomes more diffuse and forms an extended envelope. This envelope contracts over the course of its own Kelvin--Helmholtz time.
    Right panel: radius and luminosity as a function of time for the first $10^5$ yr post-disruption. The dashed lines correspond to a simplified model of a contracting polytropic envelope (see text), and the solid lines are MESA output. The rough agreement between the models supports the idea that the early ``transient'' evolution of the remnant corresponds to the contraction of an extended envelope that formed during the disruption as a result of energy deposition close to the surface of the star.}
    \label{fig:early_evolution}
\end{figure*}

In this appendix, we examine the early ($\lesssim 10^5$ yr post-disruption) evolution of the remnant's internal structure. In order to understand this evolution, we compare it to a simplified model of a contracting envelope. We will discuss the $1M\beta1$ remnant, although similar results are found for the others. 

The left panel of Figure \ref{fig:early_evolution} shows the MESA density profile of the remnant at different points in time.
As a result of the encounter, the envelope of the remnant extends to $\approx 6\, R_\odot$. The timescale for this extended envelope to contract noticeably is its Kelvin--Helmholtz time,
\begin{equation}
    t_\mathrm{KH}=\frac{G\, M_\star \, M_\mathrm{env}}{2\,R\, L}
\approx100\,\mathrm{yr}
\left(\frac{M_\star}{M_\odot}\right)
\left(\frac{M_\mathrm{env}}{10^{-2}\, M_\odot}\right)
\left(\frac{R}{5\, R_\odot}\right)^{-1}
\left(\frac{L}{300\, L_\odot}\right)^{-1}.
\end{equation}

The evolution of these extended envelopes has been studied in similar contexts; for example, \cite{Tylenda2005} reproduced the gradual decline of the stellar merger remnant V838 Mon using a model of a contracting extended envelope. Motivated by the density structure of our TDE remnant, we use that approach to study its early evolution. We refer the reader to \cite{Tylenda2005} for a detailed construction of the model, but we summarize its main properties here. 

The structure of the envelope is taken to be polytropic (for simplicity here we use a polytropic index $n = 3/2$). The energy of the envelope decreases as it contracts; the rate of change of this energy is equal (in magnitude) to its luminosity. This equality yields a differential equation for the radius of the envelope as a function of time. The effective temperature as a function of time is an input to the model; we use the MESA effective temperature. The other input to the model is the mass of the envelope, which we estimate from the MESA profiles. The mass above the radial coordinate $R_\odot$ (the approximate boundary between the interior and the extended envelope) depends on time but remains $\approx 10^{-2}\, M_\odot$. We use a constant value of $8 \times 10^{-3}\; M_\odot$ in our calculation. Finally, we use the maximum radius of the envelope $\approx 6\, R_\odot$ as the initial condition.

The right panel of Figure \ref{fig:early_evolution} compares this contracting envelope model to our MESA model of the remnant. The models agree reasonably well during the first ten thousand years post-disruption; the disagreement becomes significant when the radius of the envelope becomes comparable to the radius of the interior ($\approx R_\odot$), at which point the assumptions of the contracting envelope model begin to break down. Based on this agreement, we interpret the early evolution of the remnant as follows: The disruption deposits energy in the star. The outermost layers, which are more diffuse and loosely bound, expand significantly, then radiate the deposited energy over the course of their Kelvin--Helmholtz time. This process corresponds to the early ``transient'' phase of the bolometric luminosity.

\bibliography{bibliography}{}
\bibliographystyle{aasjournalv7}

\end{document}